# A wave function based ab initio non–equilibrium Green's function approach to charge transport

Martin Albrecht[a*], Bo Song[b], and Alexander Schnurpfeil[a,c]

[a] *Theoretical Chemistry FB08 - University Siegen - 57068 Siegen/Germany,*

[b] *Department of Physics, Beijing Technology and Business University, Beijing 100037,*

[c] *Institute for Theoretical Chemistry - University of Cologne - 50939 Köln/Germany*

## Abstract

We present a novel *ab initio* non–equilibrium approach to calculate the current across a molecular junction. The method rests on a wave function based description of the central region of the junction combined with a tight binding approximation for the electrodes in the frame of the Keldysh Green's function formalism. In addition we present an extension so as to include effects of the two–particle propagator. Our procedure is demonstrated for a dithiolbenzene molecule between silver electrodes. The full current–voltage characteristic is calculated. Specific conclusions for the contribution of correlation and two–particle effects are derived. The latter are found to contribute about 5 % to the current. The order of magnitude of the current coincides with experiments.

* Corresponding author: e-mail: m.albrecht@uni-siegen.de, Phone: +49-271 740-4024

## I. INTRODUCTION

Recent years have seen a steep rise in the broad field of nano engineering with molecular junctions being a significant part of it[1]. This has been brought about by tremendous advances in engineering techniques, which led to both unraveled reduction in size and variety[2]. The particular interest in molecular junctions is to ultimately design switches or 'transistors' on a nano scale which might be triggered by phenomena other than electrical current.

In outstanding experiments a current through a single molecule has been observed employing either a gold tip[3] as well as indeed fixing a molecule chemically well defined between two gold electrodes[4]. However, at present experiments are still notoriously difficult and hard to interpret, thus shifting weight to theoretical considerations which are expected to both create a fundamental understanding of the microscopic processes involved and provide a guidance for improved engineering.

Theoretical descriptions of the problem try to illuminate partial aspects like the role of the molecular electronic structure[5–7] or the influence of various structural conformations[8–10].

In the case of molecular junctions, LDA–based schemes were pursued by Rakshit *et al.*[11] Pantelides *et al.*[6,7] and others[5,8,10] for carbon wires and benzene–ringsystems. Most recently the DFT based augmented plane–wave method was applied to monowires by Mokrousov *et al.*[12]. Further approximations are introduced by Guitiérrez–Laliga et al. in an application to an all–carbon–system with capped nanotubes as electrodes[13]. Along those lines Fagas *et al.* analyzed the off–resonant electron transport in oligomers[14], while both discussed the switch–like behavior of a $C_{60}$ ball by means of rotation[15]. Cuniberti *et al.* focused on the role of the contacts at the junction[16,17]. An application predicting the actual current-voltage behavior of two aromatic molecules was demonstrated by Heurich *et al.*[18].

A more advanced scheme developed by Xue, Ratner and Datta sticks with these approximations, but develops a non–equilibrium formalism[19–22]. The earlier attempts were presented by Wang, Guo and Taylor.[23–25]. The principle ideas go back to Caroli *et al.* who originally focused on non–interacting systems[26]. While conceptually somewhat different, for the case of dc currents the ansatz of Cini is known to yield the same results[27]. The steady–state procedures have then been formulated as DFT schemes along the lines proposed by Lang[28].

Another set of approaches renounces completely attempts of *ab initio* calculations and resorts to empirical models[19,20,29,30]. In this frame even electron–phonon coupling can be accounted for, as reported by Gheorghe *et al.*[31]

However, possibly due to inherent shortcomings of the LDA approach, some results are far off experimental data. Calculations of Di Ventra *et al.* turned out to be off by two orders of magnitude[7,32], while different functionals turned out to lead to large fluctuations of one order of magnitude[32,33]. The reason behind this is assumed to be the fact that transmission functions obtained from static DFT approaches have resonances at the Kohn–Sham eigenenergies, which frequently do not coincide with the physical excitation energies.

In contrast, a most recent wave function based *ab initio* approach based on scattering theory came into the range of experiments for the same kind of organic systems[34]. Wave function based *ab initio* methods typically display a steeper increase in numerical cost with system size, but offer a straightforward and systematic applicability, with eigenvalues being obtainable, in principle, to any desired accuracy. Recently one of the authors developed a completely wave function based *ab initio* procedure to efficiently obtain the Green's function and related quantities for solids, polymers and molecules[35–38]. Subsequently this scheme proved valiant for the calculation of the transmission coefficient in the frame of the zero–voltage approximation to the Landauer Theory[39–41]. While these efforts where limited to the equilibrium Green's function, one of the authors recently presented a model implementation of the Keldysh non–equilibrium theory and obtained the full current–voltage characteristic for a quantum dot model[42]. The procedure relies on a formulation of Haug and Jauho[43] and replaces the equilibrium Green's function by the Keldysh Green's function which is obtained from a time integration in the complex time plane. A derivation is given by Rammer and Smith[44]. Subsequently we applied this ansatz to a dithiolethine junction[45].

The idea of the present work is to combine the accuracy of a full–fledged wave function based *ab initio* method with the Keldysh formalism as it was established for charge transport through molecular junctions. In particular we develop a scheme which allows to assess some effects of the two–particle propagator on the current. To this end we stick to the tight binding approximation of Ref.[42] as far as the electrodes are concerned, but replace the quantum dot by a realistic organic molecule which is treated together with its contacts to the electrodes on a full quantum chemical basis. An approximate treatment of the equations of motion (EOM) allow to incorporate relevant parts of the two–particle Green's function.

## II. THEORY

Local HF orbitals serve as a starting point and basis for all matrices. In terms of the local HF orbitals a model space P and excitation space Q are distinguished for the example of virtual states (the case of occupied states being completely analogous) as follows: The model space P describing the HF level comprises of the $(N+1)$–particle HF determinants $|n\rangle$, while the correlation space Q contains single (and , in principle double) excitations $|\beta\rangle$ on top of $|n\rangle$:

$$|n\rangle = c_n^\dagger|\Phi_{\mathrm{HF}}\rangle, \qquad |\beta\rangle = c_r^\dagger c_a|n\rangle, \quad c_r^\dagger c_s^\dagger c_a c_b|n\rangle \tag{1}$$

$$P = \sum_n |n\rangle\langle n|, \qquad Q = \sum_\beta |\beta\rangle\langle\beta|. \tag{2}$$

We adopt the index convention that $a, b, c, d$ and $m, n, r, s, t, u$ represent occupied and virtual HF orbitals, respectively.

### A. The one–particle Green's function

The Green's function $G_{\mathrm{nm}}(t) = -i\langle T[c_{\mathrm{n}}(0)c_{\mathrm{m}}^\dagger(t)]\rangle$, where T is the time–ordering operator and the brackets denote the average over the exact ground–state, can be obtained from Dyson's equation as:

$$G_{nm}(\epsilon) = [\epsilon 1 - \mathbf{F} - \mathbf{\Sigma}(\epsilon)]^{-1}_{\mathrm{nm}}. \tag{3}$$

Here the self energy $\Sigma_{\mathrm{nm}}(\epsilon)$ which contains the correlation effects, has been introduced and 1 represents the identity matrix. To construct the self energy the resolvent

$$\left[\epsilon 1 - \mathbf{H}^{(\mathrm{r})} + i\delta 1\right]^{-1}_{\beta;\beta'} \tag{4}$$

is needed. It can be gained from diagonalization of the Hamiltonian

$$[H^{(\mathrm{r})}]_{\beta,\beta'} = \langle\beta|H - E_0|\beta'\rangle, \tag{5}$$

where the states $|\beta\rangle, |\beta'\rangle$ are those of the correlation space Q as in Eq. (1). Here $E_0$ is the HF ground state energy while the brackets indicate the HF average. The index $^{(\mathrm{r})}$ is a reminder that we work in the 2–particle–1–hole space (2p1h).

The self energy is approximated by decomposition into a retarded and an advanced part, again indicated by the index $^{(\mathrm{r})}$ or $^{(\mathrm{a})}$, respectively. In what follows the superscript $^{(\mathrm{r})}$ will be used throughout to refer to the retarded case, while $^{(\mathrm{a})}$ is taken to denote the advanced part. Furthermore, the configuration space will be restricted to single excitations, i. e. three–body–interactions.

In the following only the construction of the retarded self energy part is given, the case of the advanced part being analogous.

The space of 2–particle 1–hole states (2p1h) is spanned by $|r,s,a\rangle = a^\dagger_{\mathrm{r}} a^\dagger_{\mathrm{s}} a_{\mathrm{a}} |\Phi_{\mathrm{HF}}\rangle$. The Hamiltonian is set up in this basis as: $[H^{\mathrm{r}}]_{\mathrm{rsa,r's'a'}} = \langle r,s,a|H - E_0|r',s',a'\rangle$ and is subsequently diagonalized.

Diagonalizing the matrix $\mathbf{H}^{(\mathrm{r})}$ results in the eigenvectors $\mathbf{S}^{(\mathrm{r})}$ and eigenvalues $\lambda^{(\mathrm{r})}$. The retarded part of the self energy is then constructed as

$$\begin{aligned}\Sigma^{(\mathrm{r})}_{\mathrm{nm}}(\epsilon) &= \sum_{\mathrm{rsa;r's'a'}} \Upsilon(\mathrm{rs;na}) \left[\epsilon 1 - \mathbf{H}^{(\mathrm{r})} + i\delta \underline{1}\right]^{-1}_{\mathrm{rsa;r's'a'}} \Upsilon(\mathrm{r's';ma'}) \\ &= \sum_{\mathrm{rsa;r's'a'}} \Upsilon(\mathrm{rs;na}) \sum_{\mathrm{q}} S^{\mathrm{r}}_{rsa;q} \frac{1}{\left(\epsilon - \lambda^{\mathrm{r}}_{\mathrm{q}} + i\delta\right)} S^{\mathrm{r}}_{q;r's'a'} \Upsilon(\mathrm{r's';ma'}). \end{aligned} \tag{6}$$

$\Upsilon$ is a shorthand for $\Upsilon(\mathrm{rs;ta}) = W_{\mathrm{rsta}} - W_{\mathrm{rsat}}$ and $W$ are the standard two–electron integrals:

$$W_{\mathrm{rsta}} := (rs|ta). \tag{7}$$

### B. The Keldysh current

This procedure is now combined with the formalism presented for a one–level quantum dot model by Yang *et al.*[42]. The current is given according to Ref.[42] by

$$J = \frac{ie}{\hbar} \int \frac{d\epsilon}{2\pi} [f_L(\epsilon) - f_R(\epsilon)] \cdot \mathbf{\Gamma}_L(\epsilon) \left[\mathbf{\Gamma}_L(\epsilon) + \mathbf{\Gamma}_R(\epsilon)\right]^{-1} \mathbf{\Gamma}_R(\epsilon) \cdot [\mathbf{G}^{(\mathrm{r})}(\epsilon) - \mathbf{G}^{(\mathrm{a})}(\epsilon)], \tag{8}$$

where the linewidth functions to the left or right of the molecular junction ($\alpha$ =L or $\alpha$ =R, respectively) have the form

$$\Gamma_{\alpha;\mathrm{mn}}(\epsilon) = H^{\alpha}_{\mathrm{m}\chi} H^{\alpha}_{\chi\mathrm{n}} 2\pi \sum_k \cos^2(k) \cdot \delta(\epsilon - \epsilon_{k,\alpha}). \tag{9}$$

In this equation the orbitals referring to the silver atoms of the central region are explicitely marked by the Greek index $\chi$, while the orbitals residing on the molecule are labeled m and n.

The overall linewidth function is just the sum:

$$\Gamma_{\mathrm{mn}} = \Gamma_{L;\mathrm{mn}} + \Gamma_{R;\mathrm{mn}}. \tag{10}$$

The energy eigenvalues $\epsilon_{k,\alpha}$ of the electrodes are given according to the tight binding approximation by

$$\epsilon_{k,\alpha} = -2tcos(k) + eV_{\alpha}, \tag{11}$$

where $V_\alpha$ is the external voltage applied to electrode $\alpha$. In fact we take $V_{\mathrm{L}} = V$ and $V_{\mathrm{R}} = 0$, where V is the external potential applied across the junction. The tight binding parameter t is taken to be half the band width of a silver chain which was reported by Springborg and Sarkar to be 5 eV[46].

The Fermi functions $f_\alpha(\epsilon)$ are defined as in Ref.[42]:

$$f_\alpha(\epsilon) = \Theta(\epsilon - D - eV_\alpha)\Theta(\mu_{\mathrm{F}} + eV_\alpha - \epsilon). \tag{12}$$

Here $D = -2t$ and $\mu_{\mathrm{F}} = 0$ are taken to be the bottom of the energy band and the Fermi energy in the leads.

Finally the Green's function including electron correlations and the linewidth function is calculated from:

$$G^{(\mathrm{r})}_{\mathrm{mn}}(\epsilon) = [\epsilon\mathbf{1} - \mathbf{F} - \mathbf{\Sigma}(\epsilon) + i\mathbf{\Gamma}(\epsilon)/2]^{-1}_{\mathrm{mn}}\,. \tag{13}$$

The extension of the quantum dot approach towards this *ab initio* treatment is visible by the replacement of the simple dot–electrode coupling parameter t' in Ref.[42] by the coupling matrices $H^{\mathrm{L}}_{\mathrm{m}\chi}$ and $H^{\mathrm{R}}_{\mathrm{m}\chi}$ in Eq. (9) as well as the inclusion of the full correlation treatment for the dithiolethine by means of the self energy matrix $\Sigma_{\mathrm{mn}}(\epsilon)$ appearing in Eq. (13).

### C. The two–particle Green's function

In order to investigate the role of the two–particle propagator we go back to the equations of motion (EOM) for the Green's functions and follow the arguments made by Jauho for the case of only one orbital level[43].

Double occupancy has been included in the original model Hamiltonian case for a quantum dot[42], but has been left out in the quantum chemical treatment on the grounds that doubly charged molecules are energetically rather expensive and should not play a major role[45]. The problem here is how the relevant parts of the two–particle Green's function can be incorporated without computing the full two–particle propagator. In the model case this was achieved by fairly simple algebra based on the form of the two–particle Green's function $g^{(2)}$ (cf. Ref.[43]) in the frame of the EOM picture. The analysis follows some basic equations which can be sketched as:

$$
\begin{aligned}
g(t-t') &= -i\left\langle T[d(t)d^{\dagger}(t')]\right\rangle && (14)\\
i\frac{\partial}{\partial t}g(t-t') &= \delta(t-t') + \epsilon g(t-t') + Ug^{(2)}(t-t'), && (15)\\
(\omega-\epsilon)g(\omega) &= 1 + Ug^{(2)}(\omega), && (16)
\end{aligned}
$$

where $d^{\dagger}$ creates a particle on the dot. The equation for $g^{(2)}$, which has been introduced with the *ad hoc* repulsion parameter U for the double occupancy of the quantum dot can be developed accordingly:

$$
\begin{aligned}
g^{(2)}(t-t') &= -i\left\langle T[d(t)n(t)d^{\dagger}(t')]\right\rangle && (17)\\
i\frac{\partial}{\partial t}g^{(2)}(t-t') &= \delta(t-t')\langle n\rangle + \epsilon g^{(2)}(t-t') + Ug^{(2)}(t-t'), && (18)\\
(\omega-\epsilon-U)g^{(2)}(\omega) &= \langle n\rangle, && (19)
\end{aligned}
$$

where $n(t) = d(t)^{\dagger}n(t)d(t)$ and the brackets denote the state average. Spin indices have been suppressed for simplicity.

We believe that a straightforward generalization of this idea for a multi–level system, (such as a realistic molecule rather than a quantum dot), is suitable for incorporation in the KELDYSH package. The idea is to allow for double occupancy on the molecule for any two virtual levels r, s, where the energy cost $U_{\mathrm{rs}}$ can be obtained on an *ab initio* level from the respective second electron affinity.

We developed an extension of the above ansatz following the EOM scheme.

For a start the general equations for the one– and two–particle Green's functions are developed. The Hamiltonian in the central region is given in terms of localized HF orbitals as:

$$
H = \sum_{\mathrm{i,j}} F_{\mathrm{ij}} c_{\mathrm{i}}^{\dagger} c_{\mathrm{j}} + \frac{1}{2}\sum_{\mathrm{ijkl}} (ij|kl)\{c_{\mathrm{i}}^{\dagger} c_{\mathrm{j}}^{\dagger} c_{\mathrm{l}} c_{\mathrm{k}}\}, \qquad (20)
$$

where we have used the notation of Lindgren and Morrison for normal ordering of operators in the last line[47,48]. The elements $(ij|kl)$ are the standard two–electron integrals (7) and $F$ is the Fock operator. The EOM for the Green's function yields for the $(N+1)$–particle space:

$$i\frac{\partial}{\partial t}G_{\mathrm{rs}}(t-t') \;=\; \delta(t-t')\left\langle c_{\mathrm{r}}(t)c^{\dagger}_{\mathrm{s}}(t')+c^{\dagger}_{\mathrm{s}}(t')c_{\mathrm{r}}(t)\right\rangle + \tag{21}$$

$$\left\langle T\left[\left[c_{\mathrm{r}}(t),H\right]c^{\dagger}_{\mathrm{s}}(t')\right]\right\rangle . \tag{22}$$

The commutator in line (22) gives rise to two parts according to the two sums contained in the Hamiltonian (20). Specifically we find:

$$\left[c_{\mathrm{r}},\sum_{\mathrm{w,x}}F_{\mathrm{wx}}c^{\dagger}_{\mathrm{w}}c_{\mathrm{x}}\right] \;=\; \sum_{\mathrm{x}}F_{\mathrm{rx}}c_{\mathrm{x}} \tag{23}$$

for the first part and

$$\left[c_{\mathrm{r}},\sum_{\mathrm{wxyz}}(wx|yz)c^{\dagger}_{\mathrm{w}}c^{\dagger}_{\mathrm{x}}c_{\mathrm{z}}c_{\mathrm{y}}\right] \;=\; \sum_{\mathrm{xyz}}\left((xr|yz)-(rx|yz)\right)c_{\mathrm{y}}c^{\dagger}_{\mathrm{x}}c_{\mathrm{z}} - \tag{24}$$

$$\sum_{\mathrm{z}}\left(\sum_{\mathrm{x}}\left((xr|xz)-(rx|xz)\right)\right)c_{\mathrm{z}}$$

for the second part. Inserting these results in Eq. (21) yields:

$$i\frac{\partial}{\partial t}G_{\mathrm{rs}}(t-t') \;=\; \delta_{\mathrm{rs}}\delta(t-t')\left\langle c_{\mathrm{r}}(t),c^{\dagger}_{\mathrm{s}}(t')\right\rangle + \tag{25}$$

$$\sum_{\mathrm{x}}\hat{F}_{\mathrm{rx}}iG_{\mathrm{xs}}(t-t')+\frac{1}{2}\sum_{\mathrm{xzy}}(xr||yz)iG^{(2)}_{\mathrm{yxzs}}(t-t'). \tag{26}$$

In the first line we have introduced the notation for the anti commutator $\{a,b\}:=ab+ba$ and in the last line the standard abbreviation $(xr||yz):=(xr|yz)-(rx|yz)$. The two–particle Green's function

$$G^{(2)}_{\mathrm{yxzs}}(t-t') \;:=\; -i\left\langle T\left[c_{\mathrm{y}}(t)c^{\dagger}_{\mathrm{x}}(t)c_{\mathrm{z}}(t)c^{\dagger}_{\mathrm{s}}(t')\right]\right\rangle \tag{27}$$

appears due to the first term on the right of Eq. 24 and couples the EOM for the one–particle Green's function to that of the two–particle Green's function. The second term on the right of Eq. (24) yields another one–particle Green's function. This term has been subsumed in a modified Fock matrix $\hat{\mathbf{F}}$ as:

$$\hat{F}_{\mathrm{rx}} \;:=\; F_{\mathrm{rx}}-\frac{1}{2}\sum_{\mathrm{z}}(zr||zx). \tag{28}$$

For the two–particle Green's function we are now seeking an approximation. We start from the respective EOM:

$$i\frac{\partial}{\partial t}G^{(2)}_{\mathrm{yxzs}}(t-t') \;=\; \delta(t-t')\left\langle\left\{c_{\mathrm{y}}(t)c^{\dagger}_{\mathrm{x}}(t)c_{\mathrm{z}}(t),c^{\dagger}_{\mathrm{s}}(t')\right\}\right\rangle + \tag{29}$$

$$i\left\langle T\left[\left[c_{\mathrm{y}}(t),H\right]c^\dagger_{\mathrm{x}}(t)c_{\mathrm{z}}(t)c^\dagger_{\mathrm{s}}(t')\right]\right\rangle +$$
$$i\left\langle T\left[c_{\mathrm{y}}(t)\left[c^\dagger_{\mathrm{x}}(t),H\right]c_{\mathrm{z}}(t)c^\dagger_{\mathrm{s}}(t')\right]\right\rangle +$$
$$i\left\langle T\left[c_{\mathrm{y}}(t)c^\dagger_{\mathrm{x}}(t)\left[c_{\mathrm{z}}(t),H\right]c^\dagger_{\mathrm{s}}(t')\right]\right\rangle .$$

The commutators can again be evaluated using Eqs. (23,24). We approximate the EOM scheme further by neglecting the resulting expressions for higher order Green's functions. The actual calculations are lengthy but without particularities. Here we just quote the result and transform to frequency space. The principle structure of the equations can be sketched as:

$$\omega G_{\mathrm{rs}}(\omega) = \delta_{\mathrm{rs}} + \sum_{\zeta}\hat{F}_{\mathrm{r}\zeta}G_{\zeta\mathrm{s}}(\omega) + \frac{1}{2}\sum_{\mathrm{y'x'z'}}\mathcal{U}_{r;y'x'z'}G^{(2)}_{\mathrm{y'x'z';s}}(\omega) \tag{30}$$

$$\omega G^{(2)}_{\mathrm{yxz;s}}(\omega) = \mathcal{N}_{\mathrm{yxz;s}} + \sum_{\mathrm{y'x'z'}}\mathcal{M}_{yxz;y'x'z'}G^{(2)}_{\mathrm{y'x'z';s}}(\omega), \tag{31}$$

where we used the shortcuts

$$\mathcal{U}_{r;yxz} := (xr||yz) \tag{32}$$

$$\mathcal{M}_{\mathrm{yxz;y'x'z'}} := \mathcal{F}_{\mathrm{yxz;y'x'z'}} - \frac{1}{2}\mathcal{Y}_{\mathrm{yxz;y'x'z'}} \tag{33}$$

$$\mathcal{F}_{\mathrm{yxz;y'x'z'}} := \delta_{\mathrm{x'x}}\delta_{\mathrm{z'z}}F_{\mathrm{yy'}} + \delta_{\mathrm{y'y}}\delta_{\mathrm{x'x}}F_{\mathrm{zz'}} - \delta_{\mathrm{y'y}}\delta_{\mathrm{z'z}}F_{\mathrm{xx'}} \tag{34}$$

$$\mathcal{Y}_{\mathrm{yxz;y'x'z'}} := 3\delta_{\mathrm{z'z}}\left((y'y|x'x) - (yx'|y'x)\right) + 3\delta_{\mathrm{z'y}}\left((zx'|y'x) - (x'z|y'x)\right) + \delta_{\mathrm{x'x}}\left((y'y||zz') + (y'z||yz')\right) \tag{35}$$

$$\mathcal{N}_{\mathrm{yxz;s}} := \left\langle\left\{c_{\mathrm{y}}c^\dagger_{\mathrm{x}}c_{\mathrm{z}}, c^\dagger_{\mathrm{s}}\right\}\right\rangle . \tag{36}$$

In fact the coupled equations (30,31) are simple matrix equations if the indices $(y,x,z)$ and $(y',x',z')$ are grouped into one triple index. The solution in matrix form is in an obvious notation:

$$\mathbf{G}(\omega) = \left[\omega\mathbf{1} - \hat{\mathbf{F}}\right]^{-1}\left(\mathbf{1} + \frac{1}{2}\mathcal{U}\left[\omega\mathbf{1} - \mathcal{F} - \frac{1}{2}\mathcal{V}\right]^{-1}\mathcal{N}\right). \tag{37}$$

Again $\mathbf{1}$ represents the identity matrix in the respective spaces.

Approximations are crucial for $N$, since a rigorous calculation would request an *a priori* knowledge of the exact ground state wave function of the full system. We apply the diagonal approximation

$$N_{\mathrm{yxzs}} \approx \delta\mathrm{ys}\delta\mathrm{xs}\delta\mathrm{zs}\left\langle n_{\mathrm{s}}\right\rangle . \tag{38}$$

In this way the averaged occupation number per energy level needs to be computed, which can be done in complete analogy to the earlier application to dithiolethine[45] or the quantum dot system[42].

In keeping with our desire to describe the case that two electrons are pushed into the central region, we keep the elements $z' = s$ and $x' = y' =: s'$ of the two particle Green's function matrix element $G_{\mathrm{y'x'z';s}}(\omega)$, so that

$$G^{(2)}_{\mathrm{y'x'z';s}}(\omega) \;\approx\; \delta_{\mathrm{z's}}\delta_{\mathrm{y'x'}}G^{(2)}_{\mathrm{x'x's;s}}(\omega). \tag{39}$$

This yields in Eq. (30) to the simplification

$$\frac{1}{2}\sum_{\mathrm{y'x'z'}}\mathcal{U}_{r;y'x'z'}G^{(2)}_{\mathrm{y'x'z';s}}(\omega) \;\approx\; \frac{1}{2}\sum_{\mathrm{s'}}\mathcal{U}_{\mathrm{s';rs's}}G^{(2)}_{\mathrm{s's's;s}}(\omega), \tag{40}$$

so that we can define an ordinary two–index matrix

$$U^{(\mathrm{s})}_{\mathrm{rs'}} \;:=\; \mathcal{U}_{\mathrm{s';rs's}} = (s'r||s's), \tag{41}$$

where in the last step we have used the definition (32).

By the same token Eq. (31) simplifies to

$$\omega G^{(2)}_{\mathrm{rrs;s}}(\omega) \;=\; \delta_{\mathrm{rs}}\left\langle n_{\mathrm{s}}\right\rangle + \sum_{\mathrm{x'}}\mathcal{M}_{rrs;x'x's'}G^{(2)}_{\mathrm{x'x's;s}}(\omega). \tag{42}$$

With the definitions (33–35) we obtain

$$\begin{aligned}\mathcal{F}_{rrs;x'x's'} &= \delta_{\mathrm{x'r}}F_{\mathrm{ss}} && (43)\\ \mathcal{V}_{rrs;x'x's'} &= 3\left((x'r|x'r) - (rx'|x'r)\right) + 3\delta\mathrm{sr}\left((sx'|x'r) - (x's|x'r)\right) + \\ &\quad \delta\mathrm{x'r}\left((x'r||ss') - (x's||rs)\right).\end{aligned}$$

This allows to introduce the ordinary two–index matrix

$$\bar{U}^{(\mathrm{s})}_{\mathrm{rx'}} \;:=\; \frac{3}{2}(x'r||x'r) + \frac{3}{2}\delta\mathrm{sr}(sx'||x'r) + \frac{1}{2}\delta\mathrm{x'r}(sr||rs). \tag{44}$$

The EOM (30, 31) finally take the form:

$$\omega G_{\mathrm{rs}}(\omega) \;=\; \delta_{\mathrm{rs}} + \sum_{\zeta}\hat{F}_{\mathrm{r}\zeta}G_{\zeta\mathrm{s}}(\omega) + \sum_{\mathrm{s'}}U^{(\mathrm{s})}_{\mathrm{rs'}}G^{(2)}_{\mathrm{s's's;s}}(\omega) \tag{45}$$

$$\omega G^{(2)}_{\mathrm{s's's;s}}(\omega) \;=\; \delta_{\mathrm{rs}}\left\langle n_{\mathrm{s}}\right\rangle + F_{\mathrm{ss}}G^{(2)}_{\mathrm{rrs;s}}(\omega) + \sum_{\mathrm{s'}}\bar{U}^{(\mathrm{s})}_{\mathrm{rs'}}G^{(2)}_{\mathrm{s's's;s}}(\omega). \tag{46}$$

To illustrate the equations we quickly show that they coincide with the ones derived by Haug and Jauho[43] for the case of a one–level quantum dot system. The dot has the orbitals $d_\uparrow$ and $d_\downarrow$. Then we have

$$U^{(\mathrm{s})}_{\mathrm{rs}'} \;\to\; \frac{1}{2}(d_\uparrow d_\downarrow|d_\uparrow d_\downarrow) =: U, \tag{47}$$

which also holds for $\bar{U}^{(\mathrm{s})}_{\mathrm{rx}'}$ in Eq. (44) and is just the Coulomb interaction $U$ felt by the one electron in the presence of the second.

Introducing $g(\omega) := G_{\mathrm{d}_\uparrow\mathrm{d}_\uparrow}(\omega)$ and defining $\epsilon_{\mathrm{d}} := F_{\mathrm{d}_\uparrow\mathrm{d}_\uparrow}$ as the energy level of the dot then yields the EOM

$$\omega g(\omega) \;=\; 1 + \epsilon_{\mathrm{d}} g(\omega) U g^{(2)}(\omega) \tag{48}$$

$$\omega g^{(2)}(\omega) \;=\; \langle n_{\mathrm{d}_\uparrow}\rangle + \epsilon_{\mathrm{d}} g^{(2)}(\omega) + U g^{(2)}(\omega), \tag{49}$$

where we used the abbreviation $g^{(2)}(\omega) := G^{(2)}_{\mathrm{d}_\uparrow\mathrm{d}_\uparrow\mathrm{d}_\downarrow;\mathrm{d}_\downarrow}(\omega)$. These are indeed the equations used in Ref.[42,43].

We now turn to a practical solution of the EOM (45, 46). Quite in analogy to the solution (37) linear algebra yields for the retarded Green's function with correlation corrections included:

$$G^{(\mathrm{r})}_{\mathrm{rs}}(\omega) \;=\; \sum_{\mathrm{s}'} \left[\omega\mathbf{1} - \hat{\mathbf{F}} - \mathbf{\Sigma}(\omega) + \frac{i}{2}\mathbf{\Gamma}(\omega)\right]^{-1}_{\mathrm{rs}'} \left[\delta_{\mathrm{s}'\mathrm{s}} + \sum_{\tilde{\mathrm{s}}} U^{(\mathrm{s})}_{\mathrm{s}'\tilde{\mathrm{s}}} Y^{-1}_{\tilde{\mathrm{s}}\mathrm{s}}(\omega)\,\langle n_{\mathrm{s}}\rangle\right], \tag{50}$$

where we introduced the shorthand

$$Y_{\tilde{\mathrm{s}}\mathrm{s}}(\omega) \;:=\; \omega\delta_{\tilde{\mathrm{s}}\mathrm{s}} - F_{\tilde{\mathrm{s}}\mathrm{s}} - \bar{U}^{(\mathrm{s})}_{\tilde{\mathrm{s}}\mathrm{s}}. \tag{51}$$

It remains to calculate $\langle n_{\mathrm{s}}\rangle$ which is done in the same way as in Ref.[42,43]. The starting point is

$$\langle n_{\mathrm{s}}\rangle \;=\; -i\int \frac{d\epsilon}{2\pi} G^{<}_{\mathrm{ss}}(\epsilon) \tag{52}$$

$$G^{<}_{\mathrm{ss}}(\omega) \;=\; \sum_{\hat{\mathrm{s}}\tilde{\mathrm{s}}} \left[G^{(\mathrm{r})}_{\mathrm{s}\hat{\mathrm{s}}}(\omega) - G^{(\mathrm{a})}_{\mathrm{s}\hat{\mathrm{s}}}(\omega)\right] \Gamma^{-1}_{\hat{\mathrm{s}}\tilde{\mathrm{s}}}(\omega)\,\left[\Gamma^{\mathrm{L}}_{\tilde{\mathrm{s}}\mathrm{s}}(\omega) f_{\mathrm{L}}(\omega) + \Gamma^{\mathrm{R}}_{\tilde{\mathrm{s}}\mathrm{s}}(\omega) f_{\mathrm{R}}(\omega)\right].$$

Inserting the expression for the retarded Green's function (50) and its advanced analogon the final result for the occupation number $\langle n_{\mathrm{s}}\rangle$ takes the form:

$$\langle n_{\mathrm{s}}\rangle \;=\; I^{\mathrm{L}}_{\mathrm{s}} + I^{\mathrm{R}}_{\mathrm{s}} + \sum_{\hat{\mathrm{s}}} \langle n_{\hat{\mathrm{s}}}\rangle \left(K^{\mathrm{L}}_{\mathrm{s}\hat{\mathrm{s}}} + K^{\mathrm{R}}_{\mathrm{s}\hat{\mathrm{s}}}\right), \tag{53}$$

which can be solved for $\langle n_{\mathrm{s}} \rangle$ once the integrals $I^{\alpha}_{\mathrm{s}}$, $K^{\alpha}_{\mathrm{s}\hat{\mathrm{s}}}$ have been computed according to:

$$I^{\alpha}_{\mathrm{s}} := -i \int \frac{d\epsilon}{2\pi} \sum_{\hat{\mathrm{s}}\tilde{\mathrm{s}}} \left[\omega \mathbf{1} - \hat{\mathbf{F}} - \mathbf{\Sigma}(\epsilon) + \frac{i}{2}\mathbf{\Gamma}(\epsilon)\right]^{-1}_{\mathrm{s}\hat{\mathrm{s}}} \Gamma^{-1}_{\hat{\mathrm{s}}\tilde{\mathrm{s}}}(\epsilon)\Gamma^{\alpha}_{\tilde{\mathrm{s}}\mathrm{s}}(\epsilon) f_{\alpha}(\epsilon) \tag{54}$$

$$K^{\alpha}_{\mathrm{s}\hat{\mathrm{s}}} := -i \int \frac{d\epsilon}{2\pi} \sum_{\mathrm{s}'\tilde{\mathrm{s}}} \left[\omega \mathbf{1} - \hat{\mathbf{F}} - \mathbf{\Sigma}(\epsilon) + \frac{i}{2}\mathbf{\Gamma}(\epsilon)\right]^{-1}_{\mathrm{s}\hat{\mathrm{s}}} U_{\mathrm{s}\mathrm{s}'} Y^{-1}_{\mathrm{s}'\hat{\mathrm{s}}}(\epsilon)\Gamma^{-1}_{\hat{\mathrm{s}}\tilde{\mathrm{s}}}(\epsilon)\Gamma^{\alpha}_{\tilde{\mathrm{s}}\mathrm{s}}(\epsilon) f_{\alpha}(\epsilon).$$

In fact Eq. (50–54) are the additional equations to be solved in order to add the two–particle effects to the current. In sum the current is then given by the integral

$$J = \frac{e}{\hbar} \sum_{\mathrm{s}\mathrm{s}'\tilde{\mathrm{s}}\hat{\mathrm{s}}} \int \frac{d\epsilon}{2\pi} [f_L(\epsilon) - f_R(\epsilon)] \cdot \Gamma^{\mathrm{L}}_{\mathrm{s}\mathrm{s}'}(\epsilon) \left[\mathbf{\Gamma}^{\mathrm{L}}(\epsilon) + \mathbf{\Gamma}^{\mathrm{R}}(\epsilon)\right]^{-1}_{\mathrm{s}'\tilde{\mathrm{s}}} \Gamma^{\mathrm{R}}_{\tilde{\mathrm{s}}\hat{\mathrm{s}}}(\epsilon) \cdot \tag{55}$$

$$\sum_{\vartheta} \left[\omega \mathbf{1} - \hat{\mathbf{F}} - \mathbf{\Sigma}(\epsilon) + \frac{i}{2}\mathbf{\Gamma}(\epsilon)\right]^{-1}_{\hat{\mathrm{s}}\vartheta} \Gamma^{-1}_{\vartheta\mathrm{s}}(\epsilon) \left[1 + \langle n_{\mathrm{s}} \rangle \sum_{\tau} U_{\hat{\mathrm{s}}\tau} Y^{-1}_{\tau\mathrm{s}}(\epsilon)\right].$$

The last factor

$$\left[1 + \langle n_{\mathrm{s}} \rangle \sum_{\tau} U_{\hat{\mathrm{s}}\tau} Y^{-1}_{\tau\mathrm{s}}(\epsilon)\right] \tag{56}$$

lends itself to an estimate of the two–particle contributions. To the one–particle effects represented by the first term '1' in Eq. (56) the effects stemming from the two–particle propagator are added as

$$\langle n_{\mathrm{s}} \rangle \sum_{\tau} U_{\hat{\mathrm{s}}\tau} Y^{-1}_{\tau\mathrm{s}}(\epsilon). \tag{57}$$

## III. RESULTS AND DISCUSSION

A sketch of the system we investigated is given in Fig. 1. A dithiolbenzene molecule is put between two silver electrodes so that the sulfur bridges bind covalently to two silver atoms, which replace the hydrogens originally present in dithiolbenzene. The silver electrodes (grey–shaded in the figure) are modeled as semi–infinite one–dimensional chains in the frame of the tight binding approximation. The central region (dashed box in the figure) is treated in a fully *ab initio* way. It contains the dithiolbenzene molecule as well as one silver atom on each side. This allows to calculate the coupling from the molecule to the silver electrodes from *ab initio* quantities as explained further below.

The central region is treated in a fully *ab initio* way. First the structure of the bare dithiolbenzene molecule is obtained by geometry optimization on the B3LYP level. Subsequently the system of the central region is constructed by replacing the hydrogen atoms of the original dithiolbenzene molecule by the silver atoms to the left and right with the Ag–S distance taken to be the average chemical value of 2.74 $\AA$. For this system canonical Hartree–Fock (HF) orbitals are calculated employing a STO–3G basis with the program package MOLPRO[49]. These orbitals are subsequently localized by means of the Pipek–Mezey option. By virtue of the program package MOLCAS a complete one– and two–electron integral transformation is performed in these orbitals, resulting in the Fock matrix $F$ and the standard two–electron integrals $W$ from Eq. (7). Finally the occupied and virtual HF orbitals $\chi$ relating to the silver electrodes are projected out from the remaining occupied as well as virtual molecular HF orbitals which are denoted by indices m and n in Eq. (8). All orbital indices are understood to also carry the spin index. The coupling matrices $H^{\mathrm{L}}_{\mathrm{m}\chi}$ and $H^{\mathrm{R}}_{\mathrm{m}\chi}$ are computed from these orbitals, where as usual L and R refer to the left and right electrode, respectively. Electronic correlations are then included by means of the program package GREENS developed by one of the authors[35–40]. To this end the self energy is constructed in a fully *ab initio* way.

For numerical evaluation of the current in Eq. (55) all matrices are transformed to the eigenstates of the linewidth matrix $\Gamma$, which are identified with the eigenchannels of the system with respect to conduction. Eigenvalues of less than 1 $\mu$Hartree were ignored.

Our result for the current–voltage characteristic is shown in Fig. 2. The overall curve displays the typical characteristics found earlier for the one–level quantum dot model[42] as well as in the case of a dithiolethine junction[45]. This is consistent with the fact that upon diagonalizing the linewidth matrix $\Gamma$ we only found one tow–fold degenerate eigenchannel. The current starts to pick up at around 1.8 $V$.

The drop–off at higher voltage finally is forced by the finite width of the silver band in the electrodes. In fact it was established in Ref.[42] that the finite band width effect sets in at about $4t$, where $t$ is the electrode band width, in our case this amounts to 10 eV, and indeed a sharp drop of the current is observed at this point in the figure.

Comparisons with experiment are delicate for various reasons. First of all we have no microscopic knowledge about the very shape of the electrodes in the central region. Secondly it is not yet clarified how precisely the molecule binds to the electrodes. Besides the possibility of a clear covalent bonding to one metal atom, intermediate bonding to two or three metal atoms at the same time where also suggested. A second source of discrepancy stems from our approximations, in particular the STO–3G basis set and the calculation with two silver atoms rather than full gold electrodes. However, it might still be instructive to compare with experiment as well as other calculations.

Quantitatively we note that the order of magnitude of some $\mu$A is in the range to be expected on the basis of experimental data on related systems. Measurements on dithiolbenzene molecules between gold electrodes suggest the same range, going up to similar values for the current[2], which was found to increase up to 0.6 $\mu$A at a voltage of 5.5 $V$. At the same voltage we find a current of 0.4 $\mu$A, so we get the same order of magnitude. In fact we compare with the very measurements reported by Reed *et al.* in Science[2] which prompted the search for theoretical descriptions. The inlet in Fig. 2 depicts the comparison. The dashed curve is taken from Fig. 4.C of Ref.[2], while the solid line shows our results.

While the order of magnitude is correct, the measurements show a current starting at about 1.0 $V$, whereas it is 1.8 $V$ in our computation. The respective plateaus resulting from the delayed onset of the current are marked in the inlet of Fig. 2 by dashed vertical lines and solid vertical lines as well as double arrows. The discrepancy in the plateau width can be attributed to the small basis set, which manifests itself in an overestimation of the fundamental gap. In particular we calculated the HF energy of the orbital responsible for the conduction both in a minimal and in a cc–pVDZ–basis set for a bare dithiolbenzene molecule and found 0.4633 $Hartree$ and 0.2025 $Hartree$, respectively. This means that basis set effects will bring down the virtual eigenenergies by about a factor two and hence the plateau will be reduced accordingly.

We now turn to an estimation of the role of the two–particle propagator. An analysis of different contributions to the current is depicted in Fig. 3.

The solid curve depicts the HF–part of the overall current, while the dashed–dotted line shows the correlation contribution. It can be clearly seen that correlations amount to about a third of the HF values and are hence not at all negligible. We also investigated the role of the two–particle propagator and found that doubly occupied molecular virtual states contribute again about 5 $\%$ of the HF results (dashed line in Fig. 3). We believe that due to the diagonal approximation Eq. (38) the role of the two–particle propagator tends to be somewhat overestimated. So it can be concluded that the two–particle Green's function has a sizeable effect on the current, but not a quantitatively crucial one.

## IV. CONCLUSIONS

In conclusion we presented a novel fully *ab initio* method to compute the current–voltage characteristic of a molecular junction. While the electrodes are still treated on the tight–binding level, a full–scale wave function based *ab initio* calculation is performed for the central region. Both the coupling of the molecular states to the electrodes as well as electronic correlations on the molecular system have been taken into account. The results enter the current both via the Green's function and the linewidth functions in the frame of the non–equilibrium Keldysh formalism. In particular we derived a scheme which allows to incorporate the effects of the two–particle propagator into the formalism. Those contributions are sizeable, but not quantitatively decisive. The order of magnitude of the current is consistent with experimental data.

We believe that this is a breakthrough which has the potential to allow for large scale applications in the future.

## V. ACKNOWLEDGMENTS

The authors are grateful for support from the German Research Foundation (DFG) in the frame of the programs SPP 1145 and AL 625/2–1.

---

[1] C. Joachim, J. K. Gimzewski, and A. Aviram, Nature (London) **408**, 541 (2000).

[2] M. A. Reed, C. Zhou, C. J. Muller, T. P. Burgin, and J. M. Tour, Science **278**, 252 (1997).

[3] C. Joachim, J. K. Gimzewski, P. R. Schlittler, and C. Chavy, Phys. Rev. Lett. **74**, 2102 (1995).

[4] C. Kergueris, J.–P. Bourgoin, S. Palacin, D. Esteve, C. Urbina, M. Magoga M, and C. Joachim, Phys. Rev. B **59**, 12505 (1999).

[5] E. G. Emberly and G. Kirczewski, Phys. Rev. B **60**, 6028 (1999).

[6] M. Di Ventra, S. T. Pantelides, and N. D. Lang, App. Phys. Lett. **76**, 3448 (2000).

[7] M. Di Ventra, S. T. Pantelides, and N. D. Lang, Phys. Rev. Lett. **84**, 979 (2000).

[8] M. P. Samanta, W. Tian, and S. Datta, Phys. Rev. B **53**, R7626 (1996).

[9] G.–L. Ingold and P. Hänggi, Chem. Phys. **281**, 199 (2002).

[10] N. D. Lang and P. Avouris, Phys. Rev. Lett. **84**, 358 (2000).

[11] T. Rakshit, G.–C. Liang, A. W. Ghosh, and S. Datta, cond–mat.: 0305695 v1 (2003).

[12] Y. Mokrousov, G. Bihlmayer, and S. Blügel, Phys. Rev. B **72**, 045402 (2005).

[13] R. Gutiérrez–Laliga, G. Fagas, K. Richter, F. Grossmann, and R. Schmidt, Europhys. Lett. **62**, 90 (2003).

[14] G. Fagas, A. Kambili and M. Elstner, *submitted to Chem. Phys. Lett. (2003).*

[15] R. Gutiérrez–Laliga, G. Fagas, G. Cuniberti, F. Grossmann, R. Schmidt and K. Richter, Phys. Rev. B **65**, 113410 (2002).

[16] G. Cuniberti, F. Großmann and R. Gutiérrez–Laliga, Advances in Sol. State Phys., Vol. **42**, Springer (2002).

[17] G. Cuniberti, G. Fagas and K. Richter, Acta Phys. Pol. B, **32**, 437 (2001).

[18] J. Heurich, J. C. Cuevas, W. Wenzel and G. Schön, Phys. Rev. Lett. **88**, 256803–1 (2002).

[19] A. W. Ghosh and S. Datta, cond–mat.: 0303630 v1 (2003).

[20] M. Paulsson, F. Zahid, and S. Datta, cond–mat.: 0208183 v1 (2002); to appear in *Nanoscience, Engineering and Technology Handbook*, William Goddard, CRC Press.

[21] Y. Xue and M. A. Ratner, Phys. Rev. B **68**, 115406 (2003).

[22] Y. Xue and M. A. Ratner, Phys. Rev. B **68**, 115407 (2003).

[23] B. Wang, J. Wang, and H. Guo, Phys. Rev. Lett. **82**, 398 (1998).

[24] J. Taylor, H. Guo, and J. Wang, Phys. Rev. B **63**, 245407 (2001).

[25] M. Brandbyge, J.–L. Mozos, P. Ordejón, J. Taylor, and K. Stokbro, Phys. Rev. B **65**, 165401 (2002).

[26] C. Caroli, R. Combescot, D. Lederer, P. Nozière, and D. Saint–James, J. Phys. C **4**, 2598 (1971).

[27] M. Cini, Phys. Rev. B **22**, 5887 (1980).

[28] N. D. Lang, Phys. Rev. B **52**, 5335 (1995).

[29] M. H. Hettler, W. Wenzel, M. R. Wegewijs, and H. Schoeller, cond–mat.: 0207483 v1 (2002).

[30] T. N. Todorov, G. A. D. Briggs, and A. P. Sutton, J. Phys.: Cond. Mat. **5**, 2389 (1993).

[31] M. Gheorghe, R. Gutiérrez–Laliga, N. Ranjan, A. Pecchia, A. Di Carlo, and G. Cuniberti, Europhysics Letters **71**, 438 (2005).

[32] F. Evers, F. Weigend, and M. Koentropp, Phys. Rev. B **69**, 235411 (2004).

[33] P. S. Krstić, D. J. Dean, X. G. Zhang, D. Keffer, Y. S. Leng, P. T. Cummings, and J. C. Wells, Comput. Mater. Sci. **28**, 321 (2003).

[34] P. Delaney and J. C. Greer, Phys. Rev. Lett. **93**, 036805 (2004).

[35] M. Albrecht and J.–I. Igarashi, J. Phys. Soc. Jap. **70**, 1035 (2001).

[36] M. Albrecht, Theor. Chem. Acc. **107**, 71 (2002).

[37] M. Albrecht and P. Fulde, Phys. Stat. Sol. b **234**, 313 (2002).

[38] M. Albrecht, Theor. Chem. Acc. **114**, 265 (2005).

[39] M. Albrecht, A. Schnurpfeil, and C. Cuniberti, Phys. Stat. Sol. b **241**, 2179 (2004).

[40] M. Albrecht, *Towards a frequency independent incremental ab initio scheme for the self energy*, accepted by Theor. Chem. Acc.; cond–mat.: 0408325 (2005).

[41] M. Albrecht, B. Song, and A. Schnurpfeil, *Charge Transport Properties of Molecular Junctions built from Dithiol Polyenes*, submitted to Theor. Chem. Acc.; cond–mat.: 0512351 (2005).

[42] K.–H. Yang, B. Song, G.–s. Tian, Y.–P. Wang, R.–S. Han, R.–Q. Han, Chin. Phys. Lett **20**, 717 (2003).

[43] H. Haug and A. P. Jauho,*Quantum Kinetics in Transport and Optics of Semiconductors*, Springer:Berlin (1996).

[44] J. Rammer and H. Smith, Rev. of Mod. Phys. **58**, 323 (1986).

[45] A. Schnurpfeil, B. Song, and M. Albrecht, *An ab initio non–equilibrium Green's function approach to charge transport: dithiolethine*, submitted to Chin. Phys. Lett. (2005); cond–mat: 0510211.

[46] M. Springborg and P. Sarkar, Phys. Rev. B **68**, 045430 (2003).

[47] I. Lindgren, and J. Morrison, *Atomic Many Body Theory*, Springer:Berlin (1985).

[48] M. Albrecht, P. Reinhardt, and J.–P. Malrieu, Theor. Chem. Acc. **100**, 241 (1998).

[49] MOLPRO, a package of *ab initio* programs designed by H.–H. Werner and P. J. Knowles, version 2002.1, R. D. Amos, A. Bernhardsson, A. Berning, P. Celani, D. L. Cooper, M. J. O. Degan, A. J. Dobbyn, F. Eckert, C. Hampel, G. Hetzer, P. J. Knowles, T. Korona, R. Lindh, A. W. Llyod, S. J. McNicholas, f. R. Manby, W. Meyer, M. E. Mura, A. Nicklass, P. Palmieri, R. M. Pitzer, R. Rauhut, M. Schütz, U. Schumann, H. Stoll, A. J. Stone, R. Tarroni, T. Thorsteinsson, and H.–J. Werner.

## Figure Captions

Fig. 1 *Sketch of the dithiolbenzene between two silver electrodes modeled as one–dimensional chains (shaded regions). The first silver atom to the left and right of the molecule replaces the hydrogens of dithiolbenzene and forms part of the central region (dashed box).*

Fig. 2 *Current–voltage characteristic for dithiolbenzene between silver electrodes as sketched in Fig. 1. The inlet shows a close–up of the same results (including negative voltage) as solid lines. For the sake of comparison measurement data are sketched as dashed lines taken from Fig. 4C in Ref.*[2].

Fig. 3 *Different contributions to the Current–voltage characteristic. HF results are shown as solid line, the correlation contributions are given by the dashed–dotted line, while the corrections due to two–particle effects are shown as dashed lines.*

# Figures

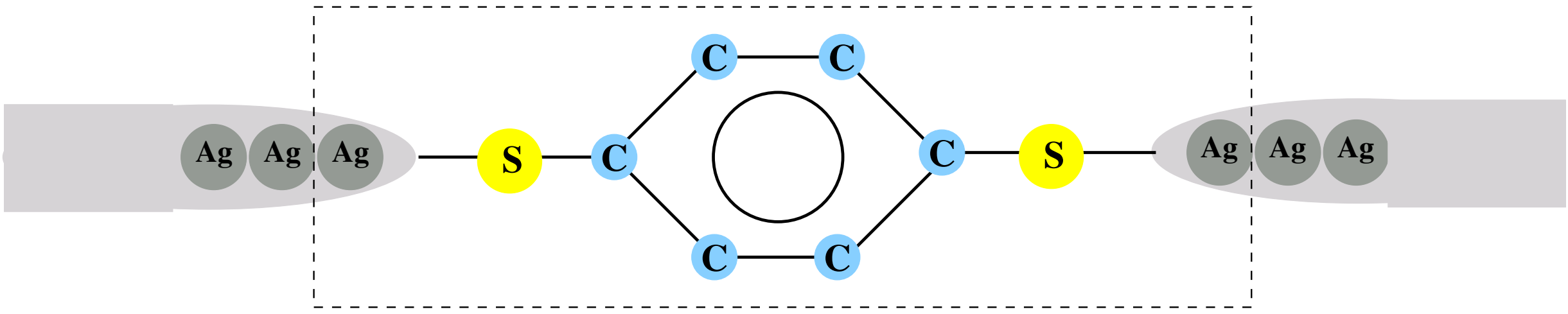

FIG. 1:

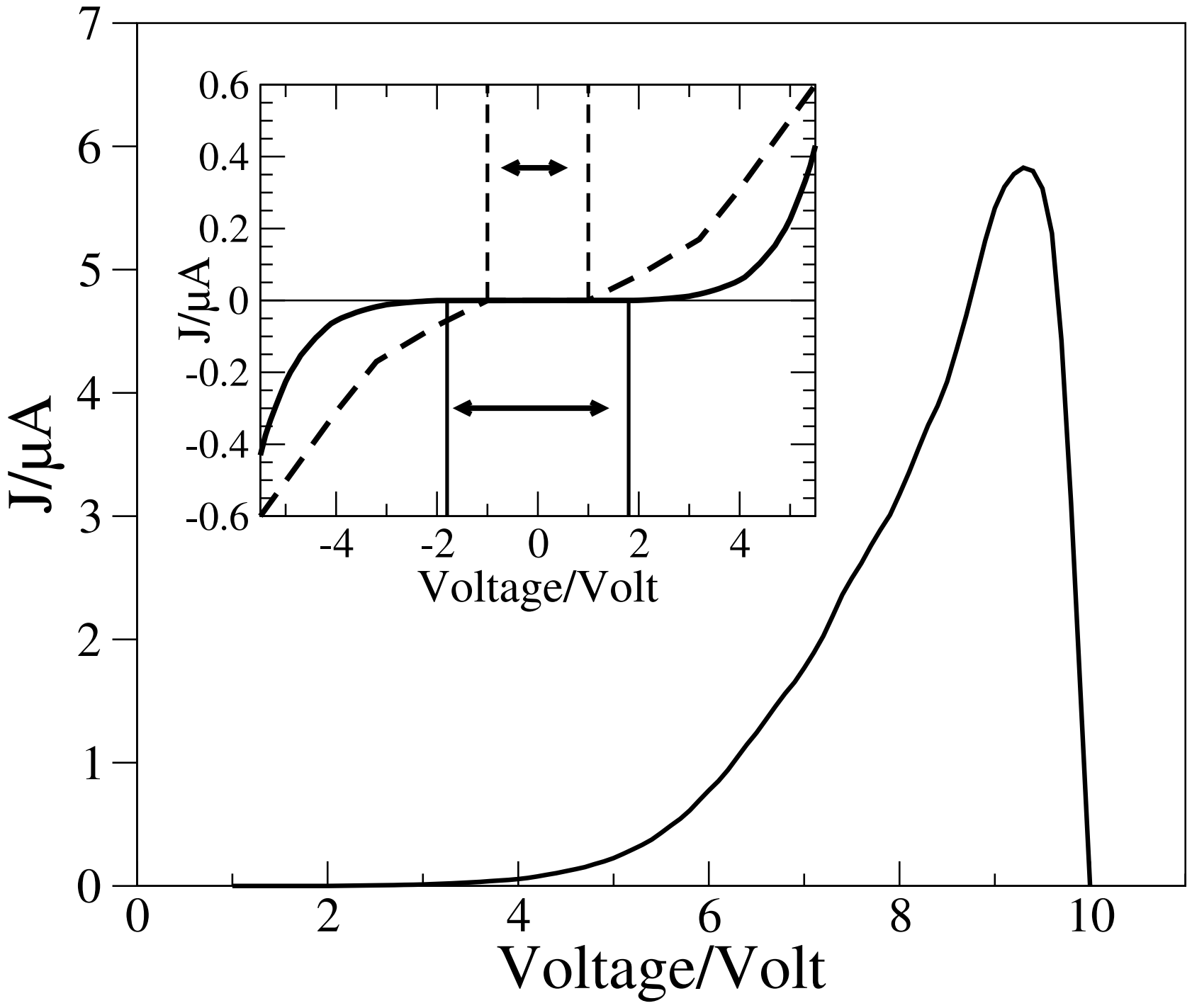

FIG. 2:

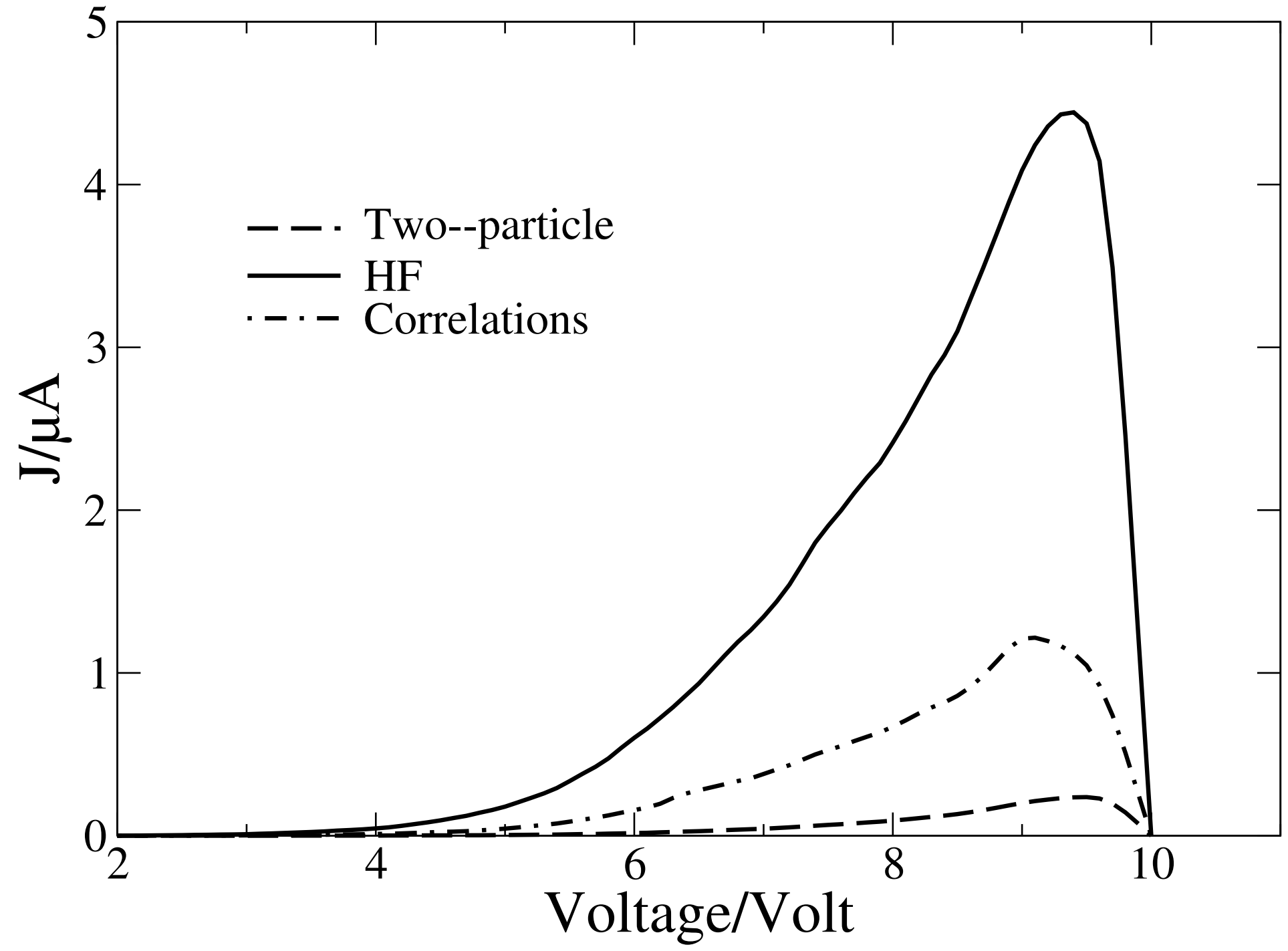

FIG. 3: